\documentclass[aps,prl,twocolumn,superscriptaddress,preprintnumbers,%
               showpacs,nofootinbib,floatfix]{revtex4}
\newcommand{\PRE}[1]{}       

\usepackage{bm}
\usepackage{graphicx,color}
\usepackage{epsfig}
\usepackage{hyperref}

\def\eslt{\not\!\!\!{E_T}}
\def\to{\rightarrow}

\def\bi{\begin{itemize}}
\def\ei{\end{itemize}}

\def\tst{\tilde t}

\def\tg{\tilde g}

\def\tw{\widetilde W}
\def\tz{\widetilde Z}

\def\alt{\lesssim}
\def\agt{\gtrsim}
\def\be{\begin{equation}}  
\def\ee{\end{equation}}  
\def\bea{\begin{eqnarray}}  
\def\eea{\end{eqnarray}}

\begin{document}

\preprint{OU-HEP-160422;\ NSF-KITP-16-041; \ UH-511-1262-16}

\title{
\PRE{\vspace*{1.5in}}
Multi-channel assault on natural supersymmetry\\
at the high luminosity LHC
\PRE{\vspace*{0.3in}}
}
\author{Howard Baer}
\affiliation{Dept. of Physics and Astronomy,
University of Oklahoma, Norman, OK, 73019, USA
\PRE{\vspace*{.1in}}
}
\author{Vernon Barger}
\affiliation{Dept. of Physics,
University of Wisconsin, Madison, WI 53706 USA and\\
Kavli Institute for Theoretical Physics,
University of California, Santa Barbara, CA 93106 USA
\PRE{\vspace*{.1in}}
}
\author{Michael Savoy}
\affiliation{Dept. of Physics and Astronomy,
University of Oklahoma, Norman, OK, 73019, USA
\PRE{\vspace*{.1in}}
}
\author{Xerxes Tata}
\affiliation{Dept. of Physics and Astronomy,
University of Hawaii, Honolulu, HI 96822, USA
\PRE{\vspace*{.1in}}
}
\date{April 25, 2016}

\begin{abstract}
\PRE{\vspace*{.1in}} Recent clarifications of naturalness in
supersymmetry robustly require the presence of four light higgsinos with mass
$\sim 100-300$ GeV while gluinos and (top)-squarks may lie in the
multi-TeV range, possibly out of LHC reach. We project the high
luminosity (300-3000 fb$^{-1}$) reach of LHC14 via gluino cascade decays
and via same-sign diboson production. We compare these to the reach for
neutralino pair production $\tz_1\tz_2$ followed by
$\tz_2\to\tz_1\ell^+\ell^-$ decay to soft dileptons which recoil
against a hard jet.  It appears that 3000 fb$^{-1}$ is just about sufficient
integrated luminosity to probe naturalness with up to 3\% fine-tuning at
the $5\sigma$ level, thus either discovering natural supersymmetry or
else ruling it out.
\end{abstract}

\pacs{12.60.-i, 95.35.+d, 14.80.Ly, 11.30.Pb}

\maketitle

The discovery \cite{atlas_h,cms_h} of a very Standard Model (SM)-like Higgs
boson at the CERN Large Hadron Collider (LHC) brings with it a puzzle.
While chiral symmetry protects fermion masses and gauge symmetry
protects gauge boson masses against large quantum corrections, no
corresponding protective symmetry exists within the SM to tame the
quadratic sensitivity of the Higgs boson mass to new physics at very
high scales. The simplest and most elegant protection, known as
supersymmetry (SUSY)\cite{wss}, relates fermions to bosons in exactly
the right way to cancel these  dangerous corrections. 
Indirect evidence for softly broken SUSY with weak scale superpartners
exists in that 1. the coupling strengths of the strong and electroweak
forces, as measured to high precision at the CERN LEP $e^+e^-$ collider
at energy scale $\sqrt{s}=m_Z$, enjoy an impressive unification at
$Q\simeq 2\times 10^{16}$ GeV when extrapolated to high energies. 
2. the top mass, measured
to be $m_t\simeq 173.2$ GeV, lies in the range required to trigger a
radiatively-induced breakdown of electroweak symmetry. Finally, 3. the
light Higgs mass $m_h$ was found to lie at $\simeq 125$ GeV, squarely
within the range required by the Minimal Supersymmetric Standard Model
(MSSM), where $m_h$ is bounded by $\alt 135$ GeV\cite{mhiggs}.
Expectations were thus heightened for the appearance of supersymmetric
matter at LHC with masses not too far above the weak scale as typified
by $m_{weak}\simeq m_{W,Z,h}\sim 100$ GeV.

In spite of these impressive 
success stories, a sense of dismay has
emerged due to the lack of evidence for direct production of
supersymmetric matter at LHC. Recent analyses of data from the first
year of LHC run 2 with $\sqrt{s}=13$ TeV $pp$ collisions and $\sim 4$
fb$^{-1}$ of integrated luminosity (just 1-2\% percent of the design
integrated luminosity sample of 300 fb$^{-1}$ even without the high
luminosity (HL) upgrade) have resulted in gluino mass bounds as high as
$m_{\tg}\agt 1.8$ TeV within the context of some simplified
models\cite{mglbound}. In addition, the rather large value of $m_h$
requires the presence of either TeV-scale highly mixed top squarks or
tens of TeV top squarks with just small left-right mixing\cite{h125}.
Such large sparticle mass values lie far beyond the classic expectations
from Barbieri-Giudice\cite{BG} (BG) naturalness  where $m_{\tg}\alt 350$ GeV 
and $m_{\tst_1}\alt 350$ were expected\cite{DG} for fine-tuning
no worse than $\sim 3\%$. The situation has led some researchers to
proclaim a crisis in physics\cite{lykken} while stimulating new,
non-supersymmetric avenues towards a solution to the gauge hierarchy
problem\cite{craig}.

An alternative response was to scrutinize the validity of the earlier
naturalness calculations\cite{comp3,MT,seige,XT}.  The simplest, most
conservative naturalness measure $\Delta_{EW}$,
proposed in Refs. \cite{ltr,rns}, is based on the well-known expression
%
%
%

\bea 
\frac{m_Z^2}{2}& =& \frac{m_{H_d}^2 + \Sigma_d^d -(m_{H_u}^2+\Sigma_u^u)\tan^2\beta}{\tan^2\beta -1} -\mu^2 
\label{eq:mzs}
\eea 
resulting from the minimization of the Higgs potential in the MSSM.
Here, $m_{H_u}^2$ and $m_{H_d}^2$ are squared soft SUSY breaking
Lagrangian terms, $\mu$ is the superpotential Higgsino mass parameter,
$\tan\beta =v_u/v_d$ is the ratio of Higgs field
vacuum-expectation-values and the $\Sigma_u^u(k)$ and $\Sigma_d^d(j)$
contain an assortment of radiative corrections, the largest of which
typically arise from the top squarks.  Expressions for the $\Sigma_u^u$
and $\Sigma_d^d$ are given in the Appendix of Ref. \cite{rns}.  The
value of $\Delta_{EW}$ compares the largest independent contribution on the
right-hand-side (RHS) of Eq.~(\ref{eq:mzs}) to the left-hand-side $m_Z^2/2$.  If
the RHS terms in Eq.~(\ref{eq:mzs}) are individually
comparable to $m_Z^2/2$, then no
unnatural fine-tunings are required to generate $m_Z=91.2$ GeV.  The
main requirements for low fine-tuning ($\Delta_{EW}\alt 30$\footnote{
The onset of fine-tuning for $\Delta_{EW}\agt 30$ is visually displayed in Ref. \cite{upper}.}) 
are the following.
\bi
\item $|\mu |\sim 100-300$
  GeV\cite{Chan:1997bi,kitnom,Barbieri:2009ew,hgsno,oldnsusy}\footnote{As in our earlier
  work, we assume that the dominant contribution to the higgsino mass is
  the superpotential $\mu$ term. A soft SUSY-breaking higgsino mass term is
  possible if there are no gauge singlets that couple to higgsinos as
  recently emphasized in Ref.~\cite{ross}.
  The authors of
  Ref.~\cite{himu} have constructed extended frameworks, with several
  additional TeV scale fields,
    to show that it 
is possible to construct natural models with heavy higgsinos.
}
(where $\mu \agt 100$ GeV is required to accommodate LEP2 limits 
from chargino pair production searches).
\item $m_{H_u}^2$ is driven radiatively to small, and not large,
negative values at the weak scale~\cite{ltr,rns}.
\item The top squark contributions to the radiative corrections
$\Sigma_u^u(\tst_{1,2})$ are minimized for TeV-scale highly mixed top
squarks\cite{ltr}.  This latter condition also lifts the Higgs mass to
$m_h\sim 125$ GeV.  For $\Delta_{EW}\alt 30$, the lighter top
squarks are bounded by $m_{\tst_1}\alt 2.5$ TeV.
\item The gluino mass which feeds into the $\Sigma_u^u(\tst_{1,2})$ via
RG contributions to the stop masses\cite{oldnsusy} 
is required to be $m_{\tg}\alt 3-4$ TeV, possibly beyond the reach of LHC.
\item First and second generation squark and slepton masses may range as
high as 5-20 TeV with little cost to naturalness\cite{rns,maren,upper}.
\ei 
SUSY models with these properties have been dubbed radiatively-driven
natural SUSY (RNS).  The presence of a high degree of fine-tuning
generally indicates a pathology or missing element in
a physical theory.

It was also found that almost all early estimates based on the BG
measure $\Delta_{BG}\equiv max_i|\partial\log m_Z^2/\partial\log p_i|$,
where the $p_i$ constitute {\em independent} fundamental parameters of
the theory, were based on application to low energy effective theories
where multiple independent soft SUSY breaking terms were introduced to
parameterize one's ignorance of hidden sector SUSY breaking. When the
underlying
correlations among soft terms (that would undoubtedly be present when
these are derived from the underlying
fundamental theory) are incorporated, it was shown that
$\Delta_{BG}\simeq \Delta_{EW}$\cite{comp3,MT,seige}. An alternative
fine-tuning measure $\Delta_{HS}$ based on large log contributions to
$m_h$ was introduced\cite{fathiggs,kitnom} which seemed to require three third
generation squarks below about 500 GeV\cite{oldnsusy}.  
This clearly ignores the possibility of correlated soft terms that
result in large cancellations in the Higgs mass.  Upon inclusion of all
independent contributions to $m_h^2$, it was found that the large log
measure also reduces to the electroweak measure\cite{comp3,seige}.

In light of these clarifications, it should not be surprising that SUSY has
not yet emerged at the LHC. Current LHC13 search limits up to $m_{\tg}\sim
1.8$ TeV\cite{mglbound} probe about half the gluino mass range allowed
by natural SUSY, while LHC13 top squark searches -- which probe to
$m_{\tst_1}\sim 750$ GeV\cite{mt1bound} -- explore much less than half
the corresponding stop mass range.  In fact, it has recently been argued
that {\it string landscape} considerations favor the higher range of
soft term values so long as the weak scale is maintained at
$m_{weak}\equiv m_{W,Z,h}\sim 100$ GeV\cite{landscape}.  The much
lighter higgsino-like charginos and neutralino $\tw_1^\pm,\
\tz_{1,2}$ can be produced at large rates at the LHC but the lightest of
these, the $\tz_1$, is assumed to comprise a portion of the dark matter
in the universe (along with {\it e.g.} axions\cite{az1}) and thus
escapes collider detection.  The heavier higgsinos have a relatively
small mass gap $m_{\tw_1,\tz_2}-m_{\tz_1}\sim 10-20$ GeV and so release
only small amounts of visible energy in their decays: thus, they are
very difficult to trigger on at the LHC much less observe above QCD
backgrounds which produce soft tracks in abundance.

\begin{figure}[tbp]
\includegraphics[height=0.35\textheight]{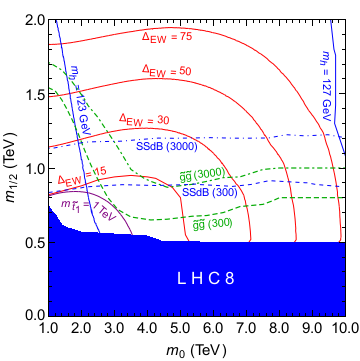}
\caption{High luminosity reach of CERN LHC for natural SUSY in the 
$m_0$ vs. $m_{1/2}$ plane for $\mu=150$ GeV, $m_A=1$ TeV, $\tan\beta =10$ and 
$A_0=-1.6 m_0$ in the NUHM2 model.
To aid the reader, we note that $m_{\tg}\sim 2.5 m_{1/2}$.
\label{fig:m0mhf}}
\end{figure}

What then are the prospects for future detection of natural SUSY at LHC?
The search for gluino pair production always figures prominently on SUSY
search menus.  In RNS SUSY, gluino pair production will be
followed by cascade decays\cite{cascade} via $\tg\to t\bar{t}\tz_i$ and
$\tg\to t\bar{b}\tw_i$ where now the lightest electroweak-inos (EWinos)
$\tw_1^\pm$ and $\tz_{1,2}$ are the light higgsino-like charginos and
neutralinos.  Gluino cascade decay events will thus be rich in $b$-jets,
typically four per event, along with isolated leptons, light quark jets
and missing $E_T$ ($\eslt$).  The $5\sigma$ reach of LHC14 with 300-3000
fb$^{-1}$ of integrated luminosity has been estimated in
Ref. \cite{andre} to extend to about $m_{\tg}\sim 1.7-2.3$ TeV within
the context of the mSUGRA/CMSSM model.  The overall LHC14 reach for
gluino pair production should be very similar for RNS SUSY as compared
to the mSUGRA model. These reach projections have been confirmed
qualitatively by CMS in Ref. \cite{cms_reach} which projects a $5\sigma$
LHC14 reach out to $m_{\tg}\sim 1950$ GeV and by Atlas\cite{atlas_reach}
which projects a $5\sigma$ LHC14 reach to $m_{\tg}\sim 2$ TeV for 300
fb$^{-1}$ and a reach to $m_{\tg}\sim 2.4$ TeV for 3000 fb$^{-1}$. The
Atlas group also quotes a $2\sigma$ (95\% CL) exclusion reach to
$m_{\tg}\sim 2.35$ TeV (2.9 TeV) for 300 (3000) fb$^{-1}$.  A
distinctive feature of RNS gluino pair production is the presence of a
dilepton mass edge in cascade decay events containing an
opposite-sign/same flavor (OS/SF) isolated dilepton pair with invariant
mass $m(\ell^+\ell^- )< m_{\tz_2}-m_{\tz_1}\sim 10-20$
GeV\cite{lhc,baris} from $\tg\to t\bar{t}\tz_2$ decay. 
The LHC14 reach for $\tg\tg$ cascade decays is
adapted from the mSUGRA study of Ref. \cite{andre} and shown in
Fig. \ref{fig:m0mhf} in the $m_0$ vs. $m_{1/2}$ plane of the
two-extra-parameter non-universal Higgs model (NUHM2)\cite{nuhm2} for
$\mu=150$ GeV, $\tan\beta =15$, $A_0=-1.6 m_0$ and $m_A=1$ TeV.  Mass
spectra were generated using Isajet 7.85\cite{isajet}.  We also show the
contours of $m_h=123$ and 127 GeV along with the current LHC8 mSUGRA
bound\cite{atlas_s,cms_s} from 20 fb$^{-1}$.

A qualitatively new SUSY signature -- same sign diboson (SSdB)
production\cite{lhcltr,lhc}-- emerges in models with light Higgsinos such as 
the RNS scenario considered in this paper. This
signal arises from wino pair production via $pp\to\tw_2\tz_4$ which is
expected to be the largest visible SUSY cross section produced at LHC14 for
$m_{\tg}\agt 1$ TeV. The winos decay mainly via $\tw_2^\pm\to
W^\pm\tz_{1,2}$ and $\tz_4\to W^\pm\tw_1^\mp$ so that half these decays
yield same sign $W$s, more of which are $W^+W^+$ events since LHC14 is a
$pp$ collider. For $W\to\ell\nu_{\ell}$ decay, then these events yield
same-sign dilepton events which are easily distinguished from SS
dilepton events arising from $\tg\tg$ production\cite{SS} in that they have
minimal accompanying jet activity-- just that arising from initial state
radiation. Heavy chargino pair production makes a subdominant but
significant contribution to the signal.

The SSdB reach along a particular RNS model line was
calculated in Ref's \cite{lhcltr,lhc}. In this work, we extend this
reach calculation into the $m_0$ vs. $m_{1/2}$ plane of
Fig. \ref{fig:m0mhf} using the hard cuts and background calculations
from Ref. \cite{lhcltr,lhc}.  The $5\sigma$ LHC14 reach with 3000
fb$^{-1}$ for SSdB production extends well beyond the $\tg\tg$ reach for
$m_0\agt 3$ TeV and extends to $m_{1/2}\simeq 1.2$ TeV corresponding to
$m_{\tg}\sim 3$ TeV. A small region with $\Delta_{EW}< 30$ extends out
beyond the LHC14 3000 fb$^{-1}$ SSdB reach. 
Confirmatory signals in the  hard- and soft- trilepton and
four-lepton channels are also possible \cite{lhc}, but the
greatest reach for
high-luminosity LHC (HL-LHC) was found to be in the SSdB channel.

While both the $\tg\tg$ and SSdB signals offer an LHC14 probe for RNS
in the $m_{1/2}$ direction, it is desirable to have some probe for
Higgsino pair production which would allow exploration in the $\mu$
direction of parameter space.  Detailed calculations of $pp\to\tz_1\tz_1
j$ production (where $j$ stands for a QCD jet arising from initial state
gluon or quark radiation) -- the so-called monojet channel-- were
performed in Ref. \cite{mono,chan1}.  There, it was found that the SUSY
signal was typically $\sim 1\%$ of QCD background which arose mainly
from $Zj$ production with $Z\to\nu\bar{\nu}$.  Thus, the monojet
signal alone does not appear to be a lucrative
discovery channel for EWino production at LHC14.

\begin{figure}[tbp]
\includegraphics[height=0.2\textheight]{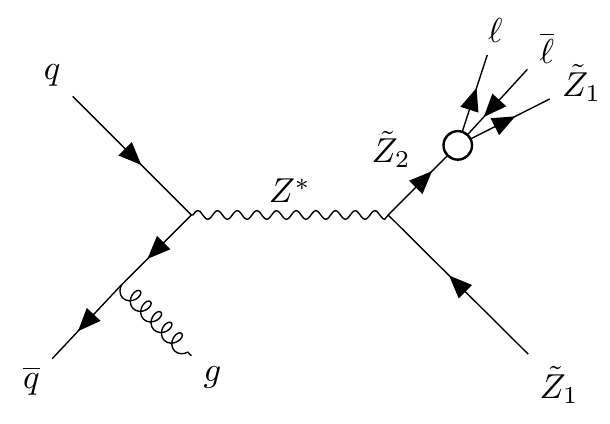}
\caption{Feynman diagram for $pp\to \tz_1\tz_2$ production
followed by $\tz_2\to\ell^+\ell^-\tz_1$ plus radiation of
a gluon jet from the initial state.
\label{fig:z1z2}}
\end{figure}

\begin{figure}[tbp]
\includegraphics[height=0.53\textheight]{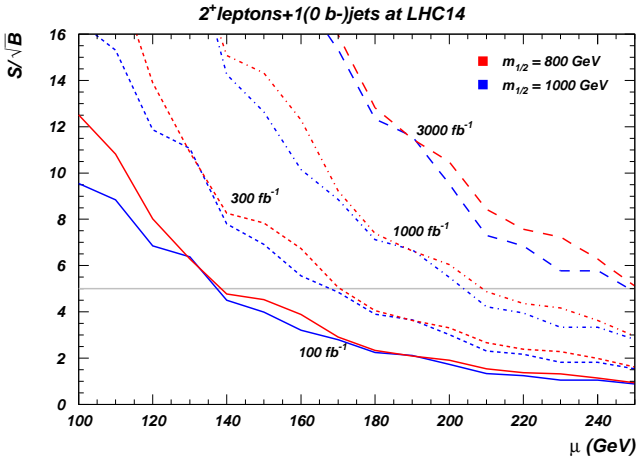}
\caption{$S/\sqrt{B}$ from HL-LHC for $pp\to\tz_1\tz_2 j$ followed by
$\tz_2\to\tz_1\ell^+\ell^-$ decay versus $\mu$, scaled for the
luminosity from Ref. \cite{dilepj}.  The blue lines are for $m_{1/2}=1000$
GeV while red lines are for $m_{1/2}=800$ GeV.
\label{fig:z1z2j}}
\end{figure}

In Ref's \cite{kribs,dilepj,chan2}, it was
suggested to look at $pp\to\tz_1\tz_2 j$ production where
$\tz_2\to\ell^+\ell^-\tz_1$.  Here, one triggers on the hard initial
state $g/q$ radiation but then requires in addition a soft OS/SF
dilepton pair with $m(\ell^+\ell^- )<m_{\tz_2}-m_{\tz_1}$ (see Fig. \ref{fig:z1z2}).
The main background occurs from SM di-tau production $pp\to Zj$ where
$Z\to\tau^+\tau^-\to \ell^+\ell^-+\eslt$.  Using $\eslt$ to reconstruct
the di-tau invariant mass, then a cut of $m^2(\tau\tau )<0$ rejected
background much more than signal\cite{dilepj}.  A small bump in the
OS/SF dilepton invariant mass distribution with $m(\ell^+\ell^-
)<m_{\tz_2}-m_{\tz_1}$ should allow determination of a signal for
sufficient integrated luminosity. We have scaled the LHC14 $5\sigma$
reach of Ref. \cite{dilepj} which is (conservatively) found to extend to
$\mu\sim 165$ GeV for 300 fb$^{-1}$ and to $\mu\sim 250$ GeV for 3000
fb$^{-1}$ of integrated luminosity (see Fig. \ref{fig:z1z2j}).

\begin{figure}[tbp]
\includegraphics[height=0.35\textheight]{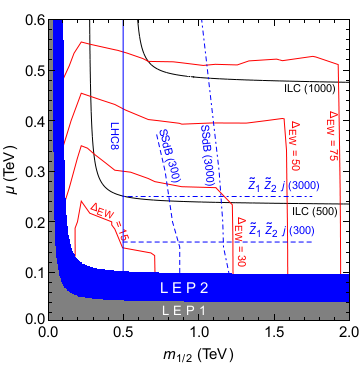}
\caption{Plot of $\Delta_{EW}$ contours (red) in the $m_{1/2}\ vs.\ \mu$
plane of NUHM2 model for $A_0=-1.6 m_0$ and $m_0=5$~TeV and $\tan\beta
=15$.  We also show the region excluded by LHC8 gluino pair searches
(left of solid blue contour), and the projected region accessible to
LHC14 searches via the SSdB channel with 300/3000 fb$^{-1}$ of
integrated luminosity (dashed/dot-dashed contours).  The LHC14 reach via
the $\tz_1\tz_2j$ channel is also shown, assuming it is insensitive to
the choice of $m_{1/2}$ in the low $\Delta_{EW}$ region of interest.  We
also show the reach of various ILC machines for higgsino pair production
(black contours).  The blue (gray) shaded region is excluded by LEP2
(LEP1) searches for chargino pair production.  To aid the reader, we
note that $m_{\tg}\simeq 2.5 m_{1/2}$.
\label{fig:mumhf}}
\end{figure}

A panoramic view of the reach of HL-LHC14 for RNS SUSY is presented in
Fig. \ref{fig:mumhf} where we show the $m_{1/2}$ vs. $\mu$ plane for
$m_0=5$ TeV, $\tan\beta =15$ $A_0=-1.6m_0$ and $m_A=1$ TeV.  The (blue)
shaded region labeled LEP2 was excluded long ago by searches for
chargino pair production at the CERN LEP2 $e^+e^-$ collider which
demands $m_{\tw_1}>103.5$ GeV for modestly large $m_{\tw_1}-m_{\tz_1}$
mass gaps.  The light Higgs mass $m_h$ lies within the $123-127$ GeV
range (the expected range of theory accuracy on our $m_h$ calculation)
throughout the entire plot.  We also show contours of $\Delta_{EW}=15$,
30, 50 and 75. The $\Delta_{EW}=30$ contour asymptotically appoaches
$\mu\sim 250$ GeV before sharply cutting off around $m_{1/2}\sim 1.2$
TeV wherein the rising top squark masses cause $\Sigma_u^u(\tst_{1,2})$
to become sufficiently large that the model becomes fine-tuned.  We also
show the present LHC8 limit for $\tg\tg$ production as the vertical line
$m_{1/2}\sim0.5$ TeV\cite{atlas_s,cms_s}.  The $5\sigma$ reach of LHC14
with 300 (3000) fb$^{-1}$ for the SSdB signal extends to $m_{1/2}\sim
0.8$ (1.2) TeV thus encompassing nearly the entire $\Delta_{EW}<30$
region (the corresponding 3000 fb$^{-1}$ LHC14 reach for $\tg\tg$
extends to $m_{1/2}\sim 1$ TeV).  We also show the 300 (3000)
fb$^{-1}$ reach of LHC14 for $\tz_1\tz_2j$ production with
$\tz_2\to\ell^+\ell^-\tz_1$ decay as dashed (dot-dashed) contours at $\mu\sim 160\
(250)$ GeV, assuming this is relatively insensitive to the precise value
of $m_{1/2}$, at least in the interesting region with low
$\Delta_{EW}$. Again, nearly the entire $\Delta_{EW}<30$ region is
covered, the exception occuring mainly at smaller $m_{1/2}\sim 0.7-1$
TeV where the $\tg\tg$ and SSdB signals should be more robust.
Throughout almost all the $\Delta_{EW}<30$ region, at least two and
sometimes all three of the RNS signals $\tg\tg$, SSdB and $\tz_1\tz_2j$
should be accessible, thus offering a degree of confirmation in multiple
signal channels. We reiterate that the SSdB signal and the soft dilepton
signal from $\tz_1\tz_2j$ production would both point to the production
of light higgsinos characteristic of RNS.  For comparison, we also show
the reach of ILC with $\sqrt{s}=0.5$ and 1 TeV. The ILC with
$\sqrt{s}\sim 0.6$ TeV should also make a decisive and complementary
search for RNS (with $\Delta_{EW} \leq 30$) via  the $e^+e^-\to
\tw_1^+\tw_1^-$ and $\tz_1\tz_2$ channels\cite{ilc}.

{\it Summary:} 

Recent clarification of electroweak naturalness points to SUSY models
containing rather light higgsinos $\sim 100-300$ GeV while gluinos and
squarks may lie in the 3-4 TeV range while maintaining naturalness at
the 3-10\% level ($\Delta_{EW}\alt 30$).  Our extension of HL-LHC SUSY
reach estimates for the planned accumulation of 3000 fb$^{-1}$ of data
displayed in Fig. \ref{fig:mumhf} shows that nearly all of natural SUSY
parameter space will be probed at the $5\sigma$ level via $\tg\tg$, SSdB
and $\tz_1\tz_2j$ searches. Signals should almost always occur in more
than one channel, thus offering strong confirmation of any
single-channel signal.  The HL-LHC 95\% CL exclusion reach typically
extends several hundred GeV further in sparticle masses. From this
vantage point, HL-LHC should either discover or exclude
radiatively-driven natural SUSY.  Further confirmation/discovery as well
as clear elucidation of the underlying scenario should occur if an
$e^+e^-$ collider with $\sqrt{s}\agt 2m(higgsino)$ such as ILC is
constructed. 
In addition, ton-scale noble liquid detectors
should detect a higgsino-WIMP signal. \cite{bbm}.

{\it Acknowledgements:} 
We thank Azar Mustafayev for providing figure 3.
This work was supported in part by the US Department of Energy, Office
of High Energy Physics. This research was supported in part by the 
National Science Foundation under Grant No. NSF PHY11-25915. 


%
\end{document}